\documentstyle[aps,twocolumn,epsf]{revtex}

\newcommand{\detau}{\partial_{\tau}}
\newcommand{\detaud}{\partial_{\tau}^{2}}

\newcommand{\nablad}{\nabla^{2}}

\newcommand{\matrice}[4]{\left(\begin{array}{cc} {#1} & {#2} \\ {#3} & {#4} \end{array} \right) }

\def\nn{\nonumber}            
\def\beq{\begin{eqnarray}}    
\def\enq{\end{eqnarray}}      
\def\ap{\left.}               
\def\at{\left(}               
\def\aq{\left[}               
\def\cp{\right.}              
\def\ct{\right)}              
\def\cq{\right]}              
\def\R{{\hbox{{\rm I}\kern-.2em\hbox{\rm R}}}}   
\def\H{{\hbox{{\rm I}\kern-.2em\hbox{\rm H}}}}   
\def\N{{\hbox{{\rm I}\kern-.2em\hbox{\rm N}}}}   
\def\C{{\ \hbox{{\rm I}\kern-.6em\hbox{\bf C}}}} 
\def\Z{{\hbox{{\rm Z}\kern-.4em\hbox{\rm Z}}}}   
\newcommand{\fr}[2]{\mbox{$\frac{#1}{#2}$}}      

\def\be{\beta}

\begin{document}


\title{Chemical potential and multiplicative anomaly \thanks{Poster presented at the 5th International
Workshop on Thermal Field Theories and Their Applications, August
10-14, 1998, Regensburg, Germany.} }

\author{Antonio Filippi \thanks{a.filippi@ic.ac.uk} }

\address{Theoretical Physics Group, Imperial
   College, \\Prince Consort Road, London SW7 2BZ, United Kingdom }

\date{September 14, 1998}

\maketitle

\begin{abstract}
The relativistic complex scalar field at finite temperature and in
presence of a net conserved charge is studied in reference to recent
developments on the multiplicative anomaly. This quantity, overlooked
until now, is computed and it is shown how it could play a role 
for this system. Other possible applications are also mentioned.
\end{abstract}

\narrowtext


\section{The charged Bose gas}
\label{sec1}
The relativistic Bose gas in presence of a net conserved charge has
been studied with some interest in recent years. The effective
potential was first obtained for the free field by Kapusta \cite{kap} 
and Haber and Weldon \cite{habwel}, 
and for the interacting field at one loop by Benson, Bernstein and
Dodelson \cite{bbd,bd}.
In recent work \cite{chemical} I presented one other possible 
approach to the inclusion of the chemical potential.

This system can be described using the standard functional integral approach
developed for thermal field theories. I will here mainly analyse the
non-interacting case, for which the
grand canonical partition function, using the most popular approach
(Method I) \cite{kap,habwel,chemical}, is
\begin{eqnarray}
Z_{\beta}(\mu)&=&\mbox{Tr}\,   e^{-\beta (H-\mu Q)}\nn \\
              &=&\int_{\phi(\tau)=\phi(\tau+\beta)}[d \phi_i]
e^{-\frac{1}{2}\int_0^\beta d\tau \int d^3x \phi_i A_{ij}\phi_j}
\:,\label{dddd}
\end{eqnarray} 
where $H$ is the Hamiltonian of the system, $Q$ the charge, $\beta$
the inverse of the temperature and $\phi_i$ the two real degrees of
freedom of the complex field. 
$A_{ij}$ is the elliptic, non-self-adjoint, matrix valued,
differential operator
\beq
\matrice{-\detaud-\nablad+m^{2}-e^2\mu^{2}}{-2ie\mu\detau}{2ie\mu\detau}{-\detaud-\nablad+m^{2}-e^2\mu^{2}}.
\enq
Notice the appearance of a ``mass term''  $m^2-e^2\mu^2$ which will eventually
give the Bose-Einstein condensation. 

We have then:
\beq
\ln Z_{\be}(\mu)= - \frac{1}{2}\ln\det \left\|
\frac{A_{ij}}{M^2}\right\|
\enq
In this case the above determinant is to be understood both as an
algebraic and a functional one. The standard procedure consists
in taking the algebraic one first \cite{kap,bbd,bd,kirtom}. 

Now, we have two possible factorizations for this
algebraic determinant:
\beq
\ln Z_{\be}(\mu)&=&\!\!-\!\frac{1}{2}\ln\det \left\|
\frac{A_{ij}}{M^2}\right\|\nn \\ 
                &=&\!\!-\!\frac{1}{2}\ln\det\!\aq
\frac{L_+}{M^2}\frac{L_-}{M^2}\cq\!\! =\!\!-\frac{1}{2}\ln\det\!\aq \frac{K_+}{M^2}\frac{K_-}{M^2}\cq
\label{o2}\!,
\enq
where:
\beq
K_{\pm}=-\nablad+m^{2}+\left(i\detau\pm ie\mu\right)^2 \: ,
\enq
\beq
L_{\pm}=-\detaud+\left(\sqrt{-\nablad+m^{2}}\pm e\mu\right)^2 .
\enq

\section{Regularization and multiplicative anomaly} 
\label{sec2}

Since these are functional determinants of
differential operators, as formal infinite products of eigenvalues they
are divergent (UV divergence) and therefore a proper regularization
scheme has to be adopted.

One of the most successful ones is the zeta-function regularization
method \cite{raysin,dowcri,haw,bcvz,libro}. 
The zeta-function regularized functional determinant of a
second order elliptic differential operator  $L$, is then defined as
\begin{eqnarray} 
\ln\det
\frac{L}{M^2} =-\zeta'(0|L)-\frac{1}{2}\zeta(0|L)\ln
M^2 \:,
\end{eqnarray} 
where $\zeta(s|L)=\mbox{Tr}\,  L^{-s}$ 
is the zeta function related to $L$, $\zeta'(0|L)$ its derivative
with respect to $s$ at zero, and $M^2$ is a renormalization scale mass. 
The fact used here is that the
analytically continued zeta-function is generally regular at $s=0$,
and thus its derivative is well defined.

Now, the determinant with which we are dealing
involves the product of two operators, as in $\log\det (AB)$. 
It is a long overlooked fact that in this case the multiplicative property
$\ln \det(AB)=\ln \det(A)+\ln \det (B)$, 
with $A$, $B$ commuting pseudo-differential operators,
does not necessarily hold.
On the contrary, an additional term $a(A,B)$,
called the {\em multiplicative anomaly} \cite{wod,konvis,kas},
may be present on the right
hand-side. Since all previous regularized computations of the effective
potential assumed the above equality, there could be additional
physical terms disregarded until now \cite{evz,efvz,nuovozerbini,bielefeld}.

It is in this light that I will reanalyze the complex scalar field at
finite temperature and chemical potential, using a properly
regularized approach and including the multiplicative
anomaly. I will mainly refer to the case of $D=4$ spacetime
dimensions. In ref. \cite{efvz}, from which this work derives, 
all the following computations have been developed for generic $D$ .

Let us for a moment assume the above equality, as in the
previous literature, so as to show the resulting inconsistencies.
I will avoid here the steps that lead to the computation of
the logarithm of the partition function, as the reader will find them in
detail in ref. \cite{efvz}.

For the first factorization $K_\pm$ we easily obtain the standard
result \cite{kap,habwel,efvz}
\begin{eqnarray}
\ln Z_{\be}(K_+,K_-)&=&\frac{\be V }{32\pi^2}\aq
m^4 (\ln \fr{m^2}{M^2}-3/2) \cq \nn \\
                  &-& V\!\! \int\!\frac{d^{3}k}{(2\pi)^{3}}\!
\ln(1\!-\!e^{-\be(\sqrt{k^{2}+m^{2}}-e\mu)})\! \nn \\ 
                  &-& V\!\! \int\!\frac{d^{3}k}{(2\pi)^{3}}\!
 \ln(1\!-\!e^{-\be(\sqrt{k^{2}+m^{2}}+e\mu)}) ,
\label{441}
\enq
where vacuum, particle and antiparticle contributions are manifest.
For the other factorization, though, the chemical potential is
associated with the momentum integral and it remains in the term
linear in $\beta$ so that we have
\beq
\ln Z_{\be}(L_+,L_-)&=&\frac{\be V }{32\pi^2}\aq
m^4 (\ln \fr{m^2}{M^2}-3/2) \cq + S(\be,\mu)\nn \\ &+& \frac{\be V }{8\pi^2} \at
\frac{e^4\mu^4}{3}-e^2\mu^2 m^2 \ct \: ,
\label{44}
\enq
where $S(\be,\mu)$ represents the standard thermal contributions as in
(\ref{441}).

In this system the importance of the multiplicative anomaly is
therefore manifest. Despite having $ \ln (K_{-}K_{+}) =\ln (L_{-}L_{+})$, 
these two options give two different results for a zeta-function regularized partition function {\em if the multiplicative
anomaly is disregarded}.

The approach taken in (\ref{o2}) was recently criticized 
 in ref. \cite{dowker}. 
According to Dowker 
the anomaly would be a result of this approach since, for generic
finite matrices, this does not correspond to taking the full algebraic
and functional determinant in one time. We replied \cite{rdowker} 
to this criticism also on general grounds but relevant
 for us now is how the above operator
is an important counter-example. The two possible factorizations can be seen
as corresponding to two possible parametrizations of the same functional
space in which we are integrating and as such linked by a
transformation in that space. Therefore the final result should clearly be
independent of the approach adopted. As I will show later, infact, the anomaly preserves this
invariance. This shows us also the delicate link between the
functional measure of the path integral and the multiplicative anomaly and
its possible physical relevance, link which clearly needs further 
investigation.

\section{The multiplicative anomaly}
\label{sec3}
Let us see what happens when the anomaly is taken in account instead.
The multiplicative anomaly \cite{wod,konvis,kas} is defined as
\begin{eqnarray} 
a_D(A,B)=\ln \det (AB)-\ln \det(A)-\ln \det (B) 
\end{eqnarray}
where the determinants of the two elliptic operators are defined by
means of the zeta-function method.
The direct computation of the unfactorized operator $\ln \det (AB)$ (to
compare with the factorized version)
is in general a too complicated task as soon as the operators are non trivial 
(see ref. \cite{evz}). Fortunately we can resort to an alternative
recipe due to Wodzicki \cite{wod}. 

For any classical pseudo-differential operator $A$ there exists
 a complete symbol $A(x,k)=e^{-ikx}Ae^{ikx}$.
This admits an asymptotic expansion for $|k| \to \infty$,
\begin{eqnarray} 
A(x,k)\sim\sum_{j=0} A_{a-j}(x,k) \:,\label{sy1}
\end{eqnarray}
 where the coefficients 
(their number is infinite) fulfil the
homogeneity property $ A_{a-j}(x,tk)= t^{a-j}A_{a-j}(x,k)$, for $t>0$
and $a$ is called the order of $A$. 
Now, Wodzicki \cite{wod} proved that for two invertible,
 self-adjoint, elliptic, commuting, pseudodifferential operators 
on a smooth compact manifold without boundaries $M_D$:
 \begin{eqnarray} 
a(A,B)=\frac{\mbox{res}\left[
(\ln(A^bB^{-a}))^2 \right]}{2ab(a+b)}=a(B,A)
\:,\label{wod3}
\end{eqnarray}
where $a >0$ and $ b> 0$ are the orders
of $A$ and $B$, respectively.
Here the quantity $\mbox{res}(A)$ is the Wodzicki non-commutative
residue. It can be computed easily using the homogeneous component
$A_{-D}(x,k)$ of order $-D$ of the complete symbol: 
\begin{eqnarray}
\mbox{res}(A)=\int_{M_D}\frac{dx}{(2\pi)^{D}}\int_{|k|=1}A_{-D}(x,k)dk
\:.\label{wod2}
\end{eqnarray}

As an example:
\beq
A(x,k)_{K_{\pm}}&=&\aq \ln \at k^2+m^2-e^2\mu^2+ i2e\mu  k_\tau
  \ct \cp \nn \\&-&\ap \ln \at k^2+m^2-e^2\mu^2- i2e\mu  k_\tau \ct \cq^2
\:.\label{903}
\enq
Remembering (\ref{sy1}),(\ref{wod2}) and (\ref{wod3}) and that the
order of our operators is 2, we have the related multiplicative anomaly as
\beq
a_4(K_+,K_-)=\frac{\be V}{8\pi^2} \aq
e^2\mu^2( m^2-\frac{e^2\mu^2}{3}) \cq
\:.\label{ex1}
\enq
The same can be done for $L_{\pm}$, obtaining a different expression
for $a_4(L_+,L_-)$.

If we now  add this two anomalies to the expressions (\ref{441}) and (\ref{44})
respectively, we obtain that the logarithm of the partition function
turns out to be the same for the two different approaches,
\beq
\ln Z_{\be}&=&
\frac{\be V }{32\pi^2}\aq
m^4 (\ln \fr{m^2}{M^2}-3/2) \cq+S(\be,\mu) \nn \\ &-&\frac{\be V}{16\pi^2} \aq
e^2\mu^2( m^2-\frac{e^2\mu^2}{3}) \cq\: .\label{441bis}
\enq

Although consistent now, our result is remarkably different from the
one in the literature where the multiplicative anomaly was
disregarded. The physical relevance of this additional term
will be discussed in the next section.

For this operators the multiplicative anomaly is vanishing for any odd
dimension $D$ and can be easily computed for any even one \cite{evz,efvz}.
In the interacting case \cite{bbd,bd},
for which the anomaly has been
computed too, it is difficult to obtain a properly
regularized expression for the rest of the partition function due to the
complexity of the operators involved \cite{efvz}.

\section{Implications}
\label{sec4}
For this system now, the effective potential in presence of external
sources, can be expressed as a function of the charge
density $\rho=\frac{1}{\be V}\frac{\partial \ln
Z_{\be}(\mu,J_i)}{\partial \mu}=\frac{<Q>}{V}$  
and the mean field ($x^2=\Phi^2 $) as
\beq
F(\be,\rho,x)&=&-\frac{1}{\be V}\ln Z_{\be}(\mu)
+\frac{\mu}{\be V} \frac{\partial \ln Z_{\be}(\mu) }{\partial
\mu}\nn \\ &+&\frac{1}{2}(m^2+e^2\mu^2)x^2 \:,
\label{v1}
\enq
\begin{eqnarray} 
\rho=\frac{1}{\be V}\frac{\partial \ln
Z_{\be}(\mu)}{\partial \mu}+ e^2\mu x^2 \:,\label{v2}
\end{eqnarray}
where the latter is an implicit expression for the chemical potential
as a function of $\rho$.
 
The physical states correspond to the minima of the effective
potential, in $\frac{\partial F}{\partial x}= x
(m^2-e^2\mu^2)=0$. We have then: 1) an unbroken phase, $x=0$,
$e\mu <m$;
2) a symmetry breaking solution, $x\neq 0$, $e\mu=\pm m$,
giving the relativistic Bose-Einstein condensation.
For our system, explicitly, the unbroken and broken phase are respectively
\beq
{\cal F}_\be &=& min F = {\cal E}_V-\frac{1}{\be V} S(\be,\mu)+\mu
\rho \nn \\  &+&\frac{1}{16\pi^2} \aq
e^2\mu^2( m^2-\frac{e^2\mu^2}{3}) \cq, \\
\rho &=&-\frac{1}{\be V}\frac{\partial S(\be,\mu)}{\partial
\mu}-\frac{e}{8\pi^2} \aq
e\mu( m^2-\frac{2 e^2\mu^2}{3}) \cq\: , 
\label{sbp111}
\enq
 where ${\cal E}_V$ is the vacuum contribution, and
\beq
{\cal F}_\be &=&{\cal E}_V-\frac{1}{\be V} S(\be,e\mu=m)+\frac{m}{e} \rho +\frac{1}{8\pi^2}\frac{m^4}{3} , \\
\rho &=&-\frac{1}{\be V}\frac{\partial S(\be,\mu)}{\partial
\mu}|_{e\mu=m}-\frac{e}{8\pi^2}\frac{m^3}{3}
+e m x^2
\:.\label{sbp1}
\enq
Ref. \cite{efvz} shows how these expressions in a generic spacetime
dimension $D$ could give some inconsistencies.
We have to remember, though, that we worked until now with regularized but
``unrenormalized'' charge density. Since it appears in the partition
function multiplied by $\mu$, any ambiguity in it will correspond to
an uncertainty in the free energy density of the kind $\mu K$.
 A physically very reasonable choice for  $K$ is given by requiring  
the symmetry to be unbroken at $T=0$, $\rho=0$. For $D=4$, $K$
will be $K=-\frac{e m^3}{24\pi^2}$. For $D=4$ only, 
this choice also removes the
multiplicative anomaly contribution to the charge density, 
so that the anomaly does not alter the broken phase 
in any respect and we get
 \begin{eqnarray}
{\cal F}_\beta &=&{\cal E}_V-\frac{1}{\beta V}
S(\beta,m)+\frac{m}{e}  \rho^R, \\ e x^2&=&
\left. \frac{1}{m}\left( \rho^R+\frac{1}{\beta V}\frac{\partial S(\beta,\mu)}{\partial \mu}\right|_{e\mu=m} \right)
\:.\label{sbp2}
\end{eqnarray}
Also the critical temperature ($x=0$, $e\mu=m$) remains unchanged with
this renormalization.
The unbroken phase is different since the anomalous term remains:
\begin{eqnarray} {\cal F}_\beta &=&{\cal E}_V-\frac{1}{\beta V}
S(\beta,\mu)+\mu \rho^R \nn \\  &-&\frac{\mu e m^3}{12 \pi^2}+\frac{1}{8
\pi^2}\left( e^2\mu^2m^2-\frac{1}{3}e^4\mu^4 \right), \\
\rho^R&=&-\frac{1}{\beta V}\frac{\partial S(\beta,\mu)}{\partial
\mu} \\ \nn &-&\frac{1}{8 \pi^2}\left( e^2\mu m^2-\frac{2}{3}e^4\mu^3 \right)
+\frac{em^3}{24\pi^2}\:. \label{sspp}
\end{eqnarray}

At ultra relativistic temperatures $T > m$ the anomalous contribution 
to the free energy is non leading,
 since the thermal contributions are proportional to  $T^4$.
At low temperatures (broken phase) we showed how the anomaly 
is reabsorbed by the renormalization of the charge, so that 
it could give relevant corrections only in a intermediate range $T\simeq m $.
Notice, finally, that the anomalous term is vanishing as
 $e \to 0$, and the correct expression of the free energy
density for the uncharged boson gas is recovered.

\section{Other systems}
\label{sec5}

Of course there are several other systems in which the
multiplicative anomaly could play a role \cite{nuovozerbini}.
A very interesting one is the non-relativistic charged scalar field,
recently reanalysed by McKenzie-Smith and Toms \cite{mcktom}. They show
how the {\em standard approach already includes 
the multiplicative anomaly term} ( which is therefore surely physical
here). On the contrary, factorising the algebraic determinant and 
forgetting the
multiplicative anomaly (as has been always done for the relativistic
case) leads to a result which does not agree with other non functional
methods.

For the relativistic field they argue that the 
multiplicative anomaly does not have new physical relevance on the ground
that the functional integral lacks a rigourous definition but they
recognize that it is crucial to obtain the correct physics
independently of the approach taken.
I have already mentioned some delicate aspects of this connection between
multiplicative anomaly and functional measure (see also
\cite{rdowker}). 

One other physical system (currently under investigation) in which
 the multiplicative anomaly could play a role is the case of fermion
mixing. Also there, as in sec. 2, there are various possible
parametrizations of the functional space involved and a transformation
(rotation) between them. Again a careful definition 
of the functional measure is necessary, this specially in the light of recent
results regarding inequivalent representations of the vacuum for field
mixing \cite{mixing}.

For other systems, though, the multiplicative anomaly could be vanishing, as it
is in odd dimensions for the one I just analyzed or for the single
fermionic field, or it could be  simply an irrelevant constant. 
For many others it should be completely reabsorvable in a 
renormalization procedure and have therefore no physical relevance,
 unless, as in the example presented in this
work, we have a presence of more than one phase, in one of which it
could survive even after the renormalization. 
There is also the problem as to whetter the multiplicative anomaly is
regularization dependent, which stirred a lively discussion lately \cite{evans,revans}.
The zeta-function method is one of a larger class of functional
regularizations called ``generalized proper-time regularizations'' \cite{sch,bal}
in which we showed the anomaly to be present \cite{bcvz,efvz}. It seems that
the answer is to be found in a proper definition of the
regularized functional determinant itself \cite{revans}.

While its relevance for zeta-function regularization is clear, the
presence of the multiplicative anomaly poses therefore many other
interesting questions about  the regularization and
renormalization  procedures, the functional integral approach itself, its
mathematical and physical definitions and its relations with alternative
 non functional approaches.  

\acknowledgments
This work has been developed in collaboration with E. Elizalde, in
Barcelona (Spain), and L. Vanzo and S. Zerbini, in Trento (Italy). My
thanks go also to R. Rivers and T. Evans for stimulating discussions.
The author wishes to
acknowledge financial support from the European Commission 
under TMR contract N. ERBFMBICT972020 and, previously, 
from the Foundation Blanceflor Boncompagni-Ludovisi, n\'ee Bildt.



\begin{references}

\bibitem{kap} J.I.~Kapusta,  Phys.~Rev. {\bf D 24}, 426 (1981).

\bibitem{habwel} H.E.~Haber and H.A.~Weldon,  Phys.~Rev.~Lett.
{\bf 46}, 1497 (1981); J.~Math.~Phys. {\bf 23}, 1852 (1982); Phys.~Rev. {\bf D
25}, 502 (1982).

\bibitem{bbd} K.~Benson, J.~Bernstein and S.~Dodelson, 
Phys.~Rev. {\bf D 44}, 2480 (1991).

\bibitem{bd} J.~Bernstein and S.~Dodelson, Phys.~Rev. Lett. {\bf 66}, 683 (1991).

\bibitem{chemical} A. Filippi,  {\em Inclusion of Chemical Potential
for Scalar Fields},   hep--ph/9703323 (1997).

\bibitem{kirtom} K.~Kirsten and D.J.~Toms,  Phys.~Lett. {\bf B
368}, 119 (1996).

\bibitem{raysin} D.B.~Ray and I.M.~Singer,  Advances in Math.
{\bf 7}, 145 (1971).

\bibitem{dowcri} J.S.~Dowker and R.~Critchley,  Phys.~Rev.
{\bf D 13}, 3224 (1976).

\bibitem{haw} S.W.~Hawking,  Commun.~Math.~Phys. {\bf 55},
133 (1977).

\bibitem{bcvz} A.A.~Bytsenko, G.~Cognola, L.~Vanzo and
S.~Zerbini,  Phys.~Rep. {\bf 266}, 1 (1996).

\bibitem{libro} E.~Elizalde, S.~D.~Odintsov, A.~Romeo, A.A.~Bytsenko
and S.~Zerbini, {\em Zeta Regularization Techniques with Applications}.
World Scientific, Singapore  (1994).

\bibitem{wod} M.~Wodzicki, {\em Non-commutative Residue Chapter I}.
In {\em Lecture notes in Mathematics}. Yu.I. Manin, editor, volume
1289,  320.   Springer-Verlag  (1987).

\bibitem{konvis} M.~Kontsevich and S.~Vishik, {\em Functional
Analysis on the Eve of the 21st Century}, volume~1,  173--197,  (1993).

\bibitem{kas} C.~Kassel,  Asterisque {\bf 177}, 199 (1989),
Sem.~Bourbaki.

\bibitem{evz} 
E.~Elizalde, L.~Vanzo and S.~Zerbini, 
Commun. Math. Phys. {\bf 194}, 613-630 (1998).

\bibitem{efvz}
E.~Elizalde, A.~Filippi, L.~Vanzo and S.~Zerbini,
Phys.~Rev. {\bf D 57}, 7430 (1998).

\bibitem{nuovozerbini}
E.~Elizalde, G.~Cognola and S.~Zerbini,
{\em Applications in physics of the multiplicative anomaly formula
involving some basic differential operators},
hep-th/9804118 (1998), to appear in Nucl. Phys. B. 

\bibitem{bielefeld}
A. Filippi,
{\em Multiplicative anomaly and finite charge density},
Proceedings of the International Workshop on QCD at Finite Baryon
Density, April 27-30, 1998, Bielefeld, Germany; hep-th/9809050 (1998).

\bibitem{dowker}
J.S.~Dowker,
{\em On the relevance of the multiplicative anomaly},
hep--th/9803200
(1998).

\bibitem{rdowker}
E.~Elizalde, A. Filippi, L. Vanzo and S. Zerbini, 
{\em Is the multiplicative anomaly relevant ?},
hep--th/980472 
(1998).

\bibitem{mcktom}
J.J.~McKenzie-Smith and D.J.~Toms,
{\em There is no new physics in the multiplicative anomaly},
hep-th/9805184
(1998).

\bibitem{mixing}
M.~Blasone, G.~Vitiello, Ann. Phys. {\bf 244}, 283, (1995);
M.~Blasone, P.A.~Henning, G.~Vitiello,
{\em Green's function approach to fermion mixing},
hep-th/9803157
(1998). 

\bibitem{evans}
T.S. Evans,
{\em Regularization schemes and the multiplicative anomaly},
hep--th/9803184
(1998).

\bibitem{revans}
E.~Elizalde, A.~Filippi, L.~Vanzo and S.~Zerbini, 
{\em Is the multiplicative anomaly dependent on the regularization~? },  
hep--th/9804071 
(1998).

\bibitem{sch} J.~Schwinger,  Phys.~Rev. {\bf 82}, 664 (1951).

\bibitem{bal} R.D.~Ball,  Phys.~Rep. {\bf 182}, 1 (1989).

\end{references}
\end{document}